\documentclass{article}
\usepackage{times}
\usepackage{amsfonts}
\usepackage[pdfmark,colorlinks]{hyperref}
\begin{document}
\title{Semiclassical Expansions, the Strong Quantum Limit, and Duality}
\author{Jos\'e M. Isidro\\
Instituto de F\'{\i}sica Corpuscular (CSIC--UVEG)\\
Apartado de Correos 22085, Valencia 46071, Spain\\
{\tt jmisidro@ific.uv.es}}

\maketitle

\begin{abstract}

\noindent
We show how to complement Feynman's exponential of the action so that it exhibits a $\mathbb{Z}_2$ duality symmetry. The latter illustrates a relativity principle for the notion of {\it quantum}\/ versus {\it classical}.

\end{abstract}

\tableofcontents

\section{Introduction}\label{pajareshijodeputa}

The motivation for this letter is taken from ref. \cite{VAFA}, section 6, from which we quote: {\it The notion of quantum versus classical is relative to which theory we measure from}. In what follows we develop a formalism that implements such a relativity principle between {\it classical}\/ and {\it quantum}. We begin by considering the exponential
\begin{equation}
{\rm exp}\left({{\rm i}\,\frac{S}{\hbar}}\right)
\label{ramallocasposo}
\end{equation}
of a certain action integral $S$. Such a function may arise, {\it e.g.}, as the integrand of a field--theoretic partition or correlation function,  as the phase of the wavefunction in a quantum--mechanical WKB approximation, etc.  The precise nature of $S$ will be immaterial for our purposes.

The generating function for the Bessel functions $J_n(x)$ of integer order $n$ (see, {\it e.g.}, ref. \cite{CH}) is $\exp\left(w(v-v^{-1})/2\right)$:
\begin{equation}
{\rm e}^{\frac{w}{2}(v-\frac{1}{v})}=\sum_{n=-\infty}^{\infty}v^n\,J_n(w), \qquad 0<\vert v\vert<\infty.
\label{pajareschupameelpiton}
\end{equation}
Setting
\begin{equation}
w=\frac{S}{\hbar}, \qquad v-v^{-1}=2{\rm i},
\label{nene}
\end{equation}
eqns. (\ref{ramallocasposo}) and (\ref{pajareschupameelpiton}) yield the expansion
\begin{equation}
{\rm e}^{{\rm i}\frac{S}{\hbar}}=\sum_{n=-\infty}^{\infty}{\rm i}^nJ_n\left(\frac{S}{\hbar}\right).
\label{eodeo}
\end{equation}

\section{Action waves}\label{ramallomekagoentuputamadre}

Let us explain the physical meaning of the expansion (\ref{eodeo}). It is an infinite sum of terms
\begin{equation}
{\rm i}^nJ_n\left(w\right),\qquad w=\frac{S}{\hbar}, \qquad n=0,\pm 1,\pm 2, \ldots,
\label{barbonmarikonketedenpordetras}
\end{equation}
where each $J_n(w)$ satisfies the Bessel equation of order $n$,
\begin{equation}
\frac{{\rm d}^2}{{\rm d}w^2}J_n(w)+\frac{1}{w}\frac{{\rm d}}{{\rm d}w}J_n(w)+\left(1-\frac{n^2}{w^2}\right)J_n(w)=0, \qquad n=0,\pm 1,\pm 2, \ldots
\label{ramallomarikonketedenpordetras}
\end{equation}
Eqn. (\ref{ramallomarikonketedenpordetras}) is closely related to a time--independent Schr\"odinger wave equation for a free nonrelativistic particle of mass $m$. 
Let ${\rm e}^{-{\rm i}Et/\hbar}$ denote the time--dependent piece of the corresponding wave function $\psi$, and consider an auxiliary plane $\mathbb{R}^2$ spanned by the dimensionless  polar coordinates $\rho, \varphi$. Then the Laplacian operator $\nabla^2$ on this auxiliary $\mathbb{R}^2$ has the radial piece \footnote{The quadratic Casimir of the Lie algebra $so(N+1)$ of rotations on $\mathbb{R}^{N+1}$ is $l(l+N-1)$.}
\begin{equation}
\frac{{\rm d}^2}{{\rm d}\rho^2}+\frac{1}{\rho}\frac{{\rm d}}{{\rm d}\rho}+\left[1-\frac{l^2}{\rho^2}\right],\qquad l=0,\pm 1,\pm 2,\ldots
\label{barbonmekagoentuputabarbacasposa}
\end{equation}
The radial variable $r$ (with dimensions of length) appearing in the Schr\"odinger equation on the plane is related to the dimensionless $\rho$ through \cite{LANDAU}
\begin{equation}
\rho=\alpha r, \qquad \alpha=\frac{1}{\hbar}\sqrt{2mE}.
\label{lex}
\end{equation}
We conclude that the $n$--th term (\ref{barbonmarikonketedenpordetras}) in the expansion (\ref{eodeo}) is a partial wave contribution to ${\rm e}^{{\rm i}\frac{S}{\hbar}}$, where the dimensionless radial coordinate $\rho$ is identified with the action $S$ as measured in units of the quantum $\hbar$, {\it i.e.},  
\begin{equation}
\rho=\frac{S}{\hbar}=w.
\label{ramalloleproso}
\end{equation}
We can identify the angular variable $\varphi$ with some function of ${\rm e}^{-{\rm i}Et/\hbar}$, because the dimensionless variable $Et/\hbar$ has a periodicity of $2\pi$. The simplest such function is a scalar multiple,
\begin{equation}
\varphi=C\,{\rm e}^{-{\rm i}Et/\hbar}.
\label{ramallokomprateunchampuantikaspa}
\end{equation}
We will see in (\ref{ramallo,piojos&grenyas&liendres}) below that $C=2\pi$. To complete the picture, the angular momentum $l\in\mathbb{Z}$ is identified with the index $n\in\mathbb{Z}$. Bearing in mind that the eigenfunctions $Y_l(\varphi)$ of the angular piece corresponding to (\ref{barbonmekagoentuputabarbacasposa}) are
\begin{equation}
Y_l(\varphi)={\rm e}^{{\rm i}l\varphi}, \qquad l=0,\pm 1,\pm 2, \ldots,
\label{ramallo,apestas-demierda}
\end{equation}
we have that Feynman's time--dependent exponential of the action becomes 
\begin{equation}
{\rm e}^{\frac{{\rm i}}{\hbar}(S-Et)}=\sum_{n=-\infty}^{\infty}{\rm i}^nJ_n\left(\frac{S}{\hbar}\right)\,\exp\left({2\pi{\rm i}n{\rm e}^{-{\rm i}Et/\hbar}}\right).
\label{ramallo,piojos&grenyas&liendres}
\end{equation}
This is an infinite sum over all possible angular momenta $n\in\mathbb{Z}$ of the auxiliary particle $m$ on the auxiliary $\mathbb{R}^2$. The choice $C=2\pi$ ensures that, when $t=0$, eqn.  (\ref{ramallo,piojos&grenyas&liendres}) reduces to eqn. (\ref{eodeo}).

Summarising, the expansion (\ref{eodeo}) decomposes Feyman's exponential of the action as an infinite sum of partial Bessel waves with circular symmetry. The latter waves propagate on an {\it auxiliary}\/ copy of $\mathbb{R}^2$; this plane is not to be confused with the physical space where the action (\ref{ramallocasposo}) is defined.  Nor are the degrees of freedom of the particle of mass $m$ propagating on $\mathbb{R}^2$ as circular waves to be confused with the degrees of freedom of the action $S$. One may call these waves on the auxiliary plane {\it action waves}. The concept of an action wave on the auxiliary plane $\mathbb{R}^2$ is extremely natural: Feynman's exponential of the action is itself an action wave, a {\it harmonic}\/ action wave, from which the usual time--dependent piece ${\rm e}^{-{\rm i}Et/\hbar}$ has been factorised.

\section{The strong quantum regime}\label{javiermas,kuandosalesdelarmario?}

In the WKB approximation to quantum mechanics \cite{LANDAU}, one substitutes ${\rm e}^{{\rm i}S/\hbar}$ for the wave function $\psi$ in the time--independent Schr\"odinger equation. To zeroth order in $\hbar$, the result equals the Hamilton--Jacobi equation of classical mechanics (for an alternative presentation of this subject see \cite{MATONE} and refs. therein). Thus, in the limit $\hbar\to 0$, Feynman's exponential of the action  provides a natural transition to classical mechanics; quantum corrections to the latter appear in powers of $\hbar$. However, is it possible to reach the strong quantum regime? By this we mean one in which $S$ becomes comparable to $\hbar$. For formal reasons it will be convenient to allow for $S$ to be smaller than $\hbar$, perhaps even much smaller. Considering $\hbar$ as a formal variable we may let the quantum of action grow as large as desired. 

Initially one may assume that the polar coordinates $\rho,\varphi$ of eqns. (\ref{lex})--(\ref{ramallokomprateunchampuantikaspa}) cover all of the auxiliary $\mathbb{R}^2$ and nothing else. However, there is no reason for $\rho,\varphi$ to be global coordinates. More generally, $\rho, \varphi$ could be local coordinates on a certain auxiliary surface $\mathbb{S}$ other than $\mathbb{R}^2$. For example, imagine that $\mathbb{S}$ is the Riemann sphere $\mathbb{CP}^1$, and let us consider the local holomorphic coordinate on $\mathbb{CP}^1$ given by
\begin{equation}
z=\rho\,{\rm e}^{{\rm i}\varphi}.
\label{barbonchupameelbolon}
\end{equation}
The circular Bessel waves (\ref{eodeo}) are outgoing with respect to the origin of coordinates $z=0$. Now the point at infinity is not covered by the coordinate (\ref{barbonchupameelbolon}). However we may reach this point by introducing the new holomorphic coordinate $\tilde z$ on $\mathbb{CP}^1$
\begin{equation}
\tilde z=-\frac{1}{z}=\tilde \rho\,{\rm e}^{{\rm i}\tilde\varphi},
\label{ramallohijoputa}
\end{equation}
where
\begin{equation}
\tilde \rho=\frac{\hbar}{S}, \qquad \tilde\varphi=-(\varphi+\pi).
\label{ramallokambiateloskalzones,almenosunavezporsemana}
\end{equation}
This leads one to the Feynman--like exponential 
\begin{equation}
{\rm exp}\left({\rm i}\,\frac{\hbar}{S}\right)
\label{ramallomariconazo}
\end{equation}
as a candidate for describing the strong quantum regime of a theory whose auxiliary surface $\mathbb{S}$ is $\mathbb{CP}^1$. Then the choice of variables in eqn. (\ref{pajareschupameelpiton})
\begin{equation}
w=\frac{\hbar}{S}, \qquad v-v^{-1}=2{\rm i}
\label{ramalloduchate,kehuelesamierda}
\end{equation}
leads to the expansion
\begin{equation}
{\rm e}^{{\rm i}\frac{\hbar}{S}}=\sum_{n=-\infty}^{\infty}{\rm i}^nJ_n\left(\frac{\hbar}{S}\right).
\label{barbonketefollen}
\end{equation}
Its physical interpretation is the same of section \ref{ramallomekagoentuputamadre}. However, {\it as seen from the viewpoint of section \ref{ramallomekagoentuputamadre}}, the Bessel waves $J_n\left(\frac{\hbar}{S}\right)$ are {\it incoming}\/ towards $z=0$, rather than outgoing. Also, their chirality is opposed to that of the outgoing waves of section \ref{ramallomekagoentuputamadre}. From the viewpoint of section \ref{ramallomekagoentuputamadre}, the semiclassical regime of (\ref{barbonketefollen}) is mapped into the strong quantum regime of (\ref{eodeo}), and viceversa. More generally, the auxiliary surface $\mathbb{S}$ need not be a sphere. It could be a Riemann surface in genus $g>0$, or no complex surface at all, perhaps just a real surface. The geometry of $\mathbb{S}$ will be dictated by the kind of duality transformations that one wishes to implement \cite{VAFA}. Obviously, theories exhibiting no dualities will have $\mathbb{S}=\mathbb{R}^2$; such is the case of standard quantum mechanics as presented, {\it e.g.}, in ref. \cite{LANDAU}.

Moreover, other decompositions of Feynman's harmonic wave (\ref{ramallocasposo}) and its close cousin (\ref{ramallomariconazo}) into a complete set of waves (of some given geometry) are also possible. For our purposes, circular waves are better suited than harmonic waves, because one can tell between {\it incoming}\/ and {\it outgoing} from the origin of coordinates $z=0$ on the auxiliary Riemann sphere $\mathbb{CP}^1$. In this way, circular waves as given by Bessel functions are useful in order to highlight a $\mathbb{Z}_2$ symmetry between the semiclassical and the strong quantum regimes. Our approach introduces an arbitrary mass $m$, that can be disposed of by setting $\alpha$ in eqn. (\ref{lex}) to any desired positive value, {\it e.g.}, $\alpha=1$. In this sense $\alpha$ plays a role analogous to that of Planck's constant: the latter sets a scale for the physical action $S$, the former sets a length scale on the auxiliary Riemann sphere $\mathbb{CP}^1$. 

Transformations between different regimes of the parameters of a given theory, leading to {\it apparently}\/ different descriptions of the same physics, are called {\it dualities}\/ \cite{VAFA}. M--theory appears as a prototype for dualities. However, as suggested in ref. \cite{VAFA} and explicitly illustrated here, one can implement dualities already within the realm of a finite number of degrees of freedom, before moving on to field theory, strings and branes. The duality illustrated in the previous example, where the auxiliary surface $\mathbb{S}$ is the Riemann sphere $\mathbb{CP}^1$, is of a very simple nature: it is just a $\mathbb{Z}_2$ duality exchanging the quantum of action $\hbar$ with the measurable action $S$. It maps the expansion (\ref{eodeo}) into the expansion (\ref{barbonketefollen}). This duality could be made manifest from the beginning in a theory whose starting point were some {\it completion}\/ of eqn. (\ref{ramallocasposo}) such as, {\it e.g.},
\begin{equation}
{\rm exp}\left({\rm i}\left(\frac{S}{\hbar}+\frac{\hbar}{S}\right)\right).
\label{ramallomarikonazo}
\end{equation}
The term ${\rm e}^{{\rm i}\frac{\hbar}{S}}$ becomes unity when $\hbar\ll S$, thus reducing to the usual exponential (\ref{ramallocasposo}). On the contrary, in the strong quantum regime, the exponential ${\rm e}^{{\rm i}\frac{\hbar}{S}}$ survives while ${\rm e}^{{\rm i}\frac{S}{\hbar}}$ tends to unity. Eqn. (\ref{ramallomarikonazo}) is manifestly selfdual, but is it  physically correct? This question can be recast as follows: can the $\mathbb{Z}_2$ duality presented above be raised to the category of a physical principle? If so, then eqn. (\ref{ramallomarikonazo}) will be physically correct. As long as only the usual exponential (\ref{ramallocasposo}) is considered, this $\mathbb{Z}_2$ duality cannot be realised explicitly. It can only be observed through an exchange of the measurable action $S$ with the quantum of action $\hbar$. In this latter case this duality is reminiscent, {\it mutatis mutandi}, of de Broglie's wave/particle duality or Bohr's complementarity principle \cite{LANDAU}.

After finishing this letter we became aware of ref. \cite{PAD}, where the notion of duality is dealt with, and some interesting applications are worked out in detail.

{\bf Acknowledgements}

It is a great pleasure to thank J. de Azc\'arraga for encouragement and support, M. Matone for interesting conversations over many years, and T. Padmanabhan for drawing attention to some important references. The author thanks Max-Planck-Institut f\"ur Gravitationsphysik (Potsdam, Germany) where this work was begun, for hospitality. This work has been partially supported by EU network MRTN--CT--2004--005104, by research grant BFM2002--03681 from Ministerio de Ciencia y Tecnolog\'{\i}a, by research grant GV2004--B--226 from Generalitat Valenciana, by EU FEDER funds, by Fundaci\'on Marina Bueno and by Deutsche Forschungsgemeinschaft.

\end{document}